\documentclass[twocolumn,showpacs,preprintnumbers,amsmath,amssymb,prb,superscriptaddress]{revtex4}

\usepackage{amsmath}
\usepackage{amssymb}
\usepackage{graphicx}
\usepackage{epsfig}
\usepackage{dcolumn}
\usepackage{textcomp}
\usepackage{bm}
\usepackage[normalem]{ulem} 
\usepackage{color} 

\def\etal{~\textit{et~al.\ }}

\newcommand{\ket}[1]{|#1\rangle}
\newcommand{\bra}[1]{\langle #1|}

\begin{document}

\title{Understanding entanglement sudden death through multipartite entanglement and quantum correlations} 

\author{Jared H. Cole}
\address{%
Institute f\"ur Theoretische Festk\"orperphysik and DFG-Center for Functional Nanostructures (CFN), 
Universit\"at Karlsruhe, 76128 Karlsruhe, Germany
}

\date{\today}



\begin{abstract}
The effect of Entanglement Sudden Death (ESD) can arise when entangling interactions convert purely bipartite entangled states into more generally entangled states.  As a result, ESD can also be seen as a function of partitioning of the system, not just of time, as the system partitioning defines different (multipartite) entanglement classes.  Computing both geometric entanglement hierarchies and the generalization of concurrence allows one to demonstrate that different methods of analysing quantum correlations provide both qualitative and quantitatively different descriptions of two commonly cited examples of ESD.  These results follow directly from the inequivalence of entanglement and quantum correlations, the later of which can exist in a state without the former.
\end{abstract}

\maketitle

\section{Introduction}
When considering the behaviour of finite state quantum systems, one usually encounters one of two characteristic types of time evolution.  The state of the system may undergo oscillatory behaviour, with characteristic frequencies given by the eigenvalues of the system Hamiltonian, which is characteristic of coherent evolution;  alternatively, the system can undergo exponential damping due to its interaction with the environment, the result being decoherence of the system.  Recently, the phenomenon of `Entanglement Sudden Death' (ESD) has been investigated, in which the bipartite entanglement between a pair of two-state systems displays neither behaviour, but instead disappears at some \emph{finite} time due to decoherence acting on each system independently.  In this paper we investigate the link between ESD and multipartite entanglement.  In pursuing this link, we are able to comment on entanglement invariants and ESD, show the equivalence of the two canonical examples of ESD, investigate the link between entanglement and correlations and show that ESD can also occur as a function of subsystem partitioning, rather than time.  

The fact that a partially mixed state can evolve from finite to zero entanglement without becoming completely mixed has been known for some time~\cite{Zyczkowski:01, Diosi:03,Dodd:04}.  This concept has been distilled into concrete examples and is referred to as the sudden death (or birth) of entanglement\cite{Yu:04, Yu:06, Yoenac:06, FrancaSantos:06,TerraCunha:07,Yu:07,Almeida:07,Ficek:08,Lopez:08,Aolita:08, Salles:08,Man:09,Yu:09}.  There are two canonical systems in which entanglement sudden death is discussed.  The most striking example consists of a pair of two-level atoms, initially in some entangled state, undergoing spontaneous emission into the environment.  These atoms have no interaction between them but are initialized in the (partially entangled) state~\footnote{This initial state comes from Refs.~\onlinecite{FrancaSantos:06,Lopez:08} but displays all the effects discussed in the mixed state version~\cite{Yu:04}.}
\begin{equation}\label{eq:esdexample1}
\ket{\phi} = \cos\theta \ket{gg} + \sin\theta \ket{ee}
\end{equation} 
where $\theta$ parameterizes the fraction of ground ($g$) and excited ($e$) states of the atoms.  The atoms undergo independent spontaneous emission, modelled using Lindblad~\cite{Nielsen:00,Gardiner:91} or Kraus~\cite{Nielsen:00,Kraus:83} operators, or similar formalism.  For $\theta \le \pi/4$, the entanglement between the atoms decays exponentially with a decay rate equal to $\Gamma$.  In contrast, when $\theta > \pi/4$, the entanglement disappears at a finite time~\cite{FrancaSantos:06,Lopez:08}, given by $t_d=-\ln[1-\cot\theta]/\Gamma$.  A state which displays this ESD is referred to as a `fragile' state~\cite{Yu:04}.

A second example system in which entanglement sudden birth and death is observed~\cite{Yoenac:06, Yu:07} is the Jaynes--Cummings (JC) system, which is well known from the study of strongly coupled atoms in cavities~\cite{Scully:97}.  Consider a pair of two-level atoms, each contained in a strongly coupled cavity where neither the atoms nor the cavities interact.  In this case the system undergoes cycles of entanglement death and birth, even though the evolution of the system is entirely coherent.  

\begin{figure}[tb!]
\centering{\includegraphics[width=8cm]{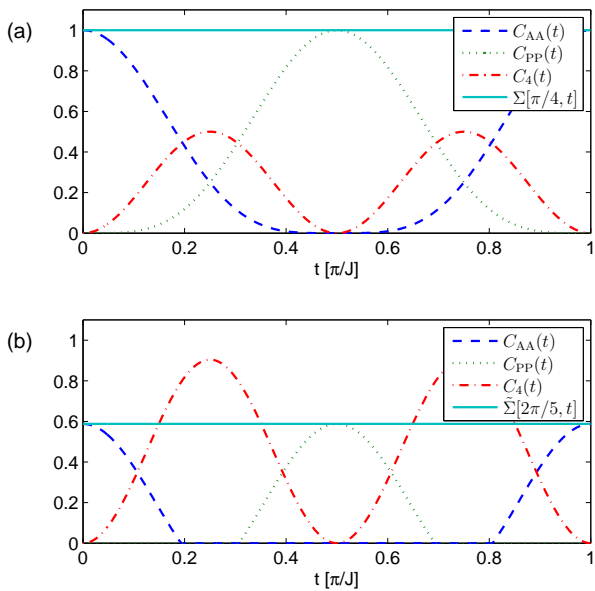}} 
\caption{Evolution of entanglement for the JC model.  Part~(a), $\ket{\psi(\pi/4, t)}$, shows the entanglement reaching zero periodically for both atoms (dash line) and photons (dotted line), even though there is no interaction \emph{between} the atoms or the photons. Part~(b) demonstrates the same evolution for the state $\ket{\psi(2\pi/5, t)}$.  The concurrences between atoms and photons is zero for finite time periods, even though the evolution is entirely coherent.  The four-particle concurrence (dash-dotted line) can be used to define an invariant (solid line).\label{fig:jcevol}}
\end{figure}

Throughout this paper we use the basis $\ket{\psi}=\ket{\rm{at}_{(1)}}\otimes \ket{\rm{ph}_{(1)}}\otimes \ket{\rm{at}_{(2)}}\otimes \ket{\rm{ph}_{(2)}}$ where $\ket{\rm{at}}$ consists of the ground and excited state of the atom, $\ket{g}$ and $\ket{e}$, and the photon mode $\ket{\rm{ph}}$ can be occupied $\ket{1}$ or unoccupied $\ket{0}$.  The atoms are initialized in the state
\begin{equation}\label{eq:esdexample2}
\ket{\psi}=\cos \theta \ket{g\, 0\, g\, 0} + \sin \theta \ket{e\, 0\, e\, 0}
\end{equation}
where $\theta$ parameterizes the fragility of the state.  The evolution of each atom-cavity pair is governed by the standard Jaynes--Cummings Hamiltonian,
\begin{equation}
H_{\rm{JC}}=\mathcal{E} \sigma_+ \sigma_- + \omega a^{\dag} a + J(a^\dag \sigma_- + a \sigma_+),
\end{equation}
where $\sigma$ acts on the atom, $a$ on the photon mode and $J$ is the coupling strength in units such that $\hbar=1$.  We take $\mathcal{E}=\omega$ and assume that the atoms and photons form a closed system and therefore ignore other decohering effects.  Given this initial state, Eq.~(\ref{eq:esdexample2}), we can write down explicitly the time evolution, 
\begin{eqnarray}\label{eq:JCsolution}
\ket{\psi(\theta, t)} & = & \cos \theta \ket{g\, 0\, g\, 0} \nonumber \\
& & + \sin \theta \bigg[ \cos^2(J t)\ket{e\, 0\, e\, 0} -  \sin^2(J t)\ket{g\, 1\, g\, 1} \nonumber \\
&  & \left. -\frac{i \sin(2 J t)}{2} \left(\ket{g\, 1\, e\, 0} + \ket{e\, 0\, g\, 1} \right)  \right].
\end{eqnarray}
The population of the system oscillates between the atom and photon degrees of freedom with a period given by $J$.  The entanglement between the atoms or the cavity photon modes also oscillates as the populations oscillate~\cite{Yoenac:06}.  For certain values of $\theta$, the entanglement can disappear for an appreciable fraction of the cycle before returning.  Fig.~\ref{fig:jcevol}(a) shows the concurrence~\cite{Wootters:98} between the atoms  and the photons as a function of time.  The interesting region is halfway through one of these cycles where both the atoms and the photons carry some entanglement, but the sum of this entanglement does not add up to that initially contained in the atoms.  Fig.~\ref{fig:jcevol}(b) shows the various concurrences for $\theta=2\pi/5$.  Note the extended periods where both the concurrence between atoms and photons is zero, even though the system is completely coherent.


We now investigate the properties of the JC example, the ESD behaviour of which has been extensively studied\cite{Yoenac:06, Yu:07}, with particular focus on the \emph{multipartite entanglement} characteristics.  The analysis of the movement of entanglement between groups of spins has been extensively studied in the context of entanglement transfer~\cite{McHugh:06,Cavalcanti:06, Serafini:06, Paternostro:04} but here we specifically make connections between these ideas and those of ESD.  Our results will additionally prove to be applicable to the first (spontaneous emission) example.  

We first ask a very important question, what does the \emph{four particle entanglement} of the system look like as a function of time?  A system comprising three qubits can be entangled in two non-equivalent classes~\cite{Duer:00}, the canonical $W$ and $\rm{GHZ}$ classes, which cannot be interconverted using only stochastic local operations and classical communication (SLOCC).   In the case of four particles, there are several non-equivalent classes under SLOCC~\cite{Verstraete:02,Lamata:06, Lamata:07,Osterloh:05, Osterloh:06}, although their exact number and boundaries are still a source of some debate.

We will start by classifying the case of $\theta=\pi/4$, which displays transitory ESD, see Fig.~\ref{fig:jcevol}(a).  The initial state $\ket{\psi(\pi/4, 0)}$ consists of 1 EPR pair shared between the atoms.  After one half cycle ($t=\pi/2J$), the EPR pair is now swapped to the photons.  These states, $\ket{\psi(\pi/4, 0)}$ and $\ket{\psi(\pi/4, \pi/2J)}$ are equivalent under exchange of the atom/photon degrees of freedom and belong to the class $L_{\rm{a_2b_2}}$ according to Verstraete \etal~\cite{Verstraete:02} or the degenerate class $0_2 0_4 \Psi_{13}$ according to Lamata~\etal~\cite{Lamata:06, Lamata:07}.  The more interesting state between these two is $\ket{\psi(\pi/4, \pi/4J)}$, which demonstrates a reduction in the total bipartite entanglement and corresponds to class $L_{\rm{abc_2}}$ or $span\{0_1 \Psi_{23},\rm{GHZ}\}$ respectively.  We see immediately that the system is changing (global) entanglement class as it evolves in time through the interaction between local pairs of particles (in this case atom--photon pairs).
The different four-particle entanglement classes have different degrees of `visibility' when measured with a bipartite measure.  
This provides us with an alternative interpretation of ESD, as the Hamiltonian evolution between multipartite entanglement classes with different visibility to bipartite entanglement.



\section{Spontaneous emission}
We now turn our attention to the first cited example of ESD, that of two entangled atoms undergoing spontaneous emission.  For this example, we use the Weisskopf-Wigner (WW) theory of spontaneous emission which involves a coherent description of an atom interaction with an infinite number of vacuum modes~\cite{Weisskopf:30,Scully:97}.  For simplicity, we will consider only modes which differ in frequency and ignore the angular and dipole element dependence.  
Using this simplified model~\cite{Lopez:08}, the Hamiltonian of a single atom/photon system is
\begin{equation}
	H_{\rm{WW}} = \mathcal{E} \sigma_+ \sigma_- + \sum_k^N \omega_k a_k^{\dag} a_k + \sum_k^N J_k(a_k^\dag \sigma_- + a_k \sigma_+)
\end{equation}
for an atom coupled to $N$ vacuum modes.

The solution to the Schr\"odinger equation under this Hamiltonian takes the form
\begin{equation}\label{eq:WWstate}
\ket{\Phi_t} = \xi(t)\ket{e}\ket{\mathbf{\overline{0}}} + \sum_{k=1}^N \lambda_k(t) \ket{g} \ket{1_k},
\end{equation}
where $\ket{\mathbf{\overline{0}}}$ is the empty vacuum state (assuming zero temperature environment).  In the limit $N \rightarrow \infty$, $\xi(t)=\exp(-\Gamma t)$, giving the usual exponential decay for an atom emitting into the vacuum.  We then define a collective mode $\ket{\gamma}$ such that $\ket{\Phi_t} = \xi(t)\ket{e}\ket{0} + \chi(t)\ket{g}\ket{\gamma}$.  A direct mapping between the WW and JC examples is given by setting $\xi(t)=\cos(J t)$ and $\chi(t)=-i\sin(J t)$.  

\begin{figure} [tb!]
\centering{\includegraphics[width=8cm]{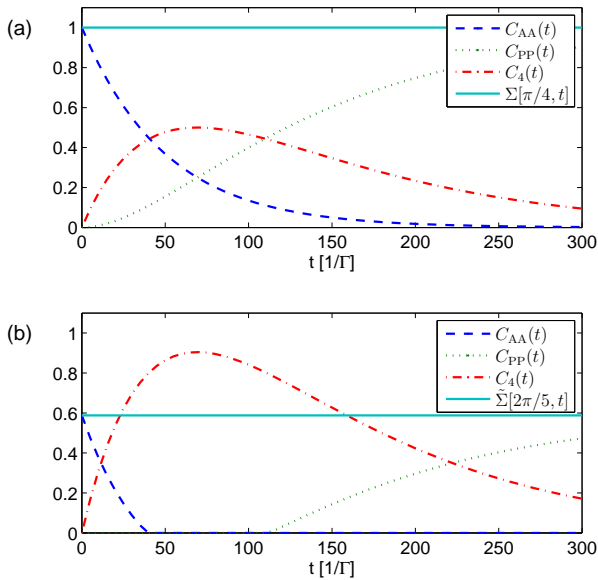}} 
\caption{Evolution of entanglement for the WW model, with identical labelling to Fig.~\ref{fig:jcevol}.  Part~(a), $\ket{\psi(\pi/4, t)}$, shows the entanglement between the atoms decaying over time, while the entanglement between the emitted photons increases, as the excitation is swapped into the photon modes. Part~(b) demonstrates the same evolution for the state $\ket{\psi(2\pi/5, t)}$, where the entanglement between atoms disappears at $t_d$ and the entanglement between photons appears at $t_b$.  The 4-partite concurrence rises and falls, reflecting the distribution of entanglement between all the degrees of freedom.  As in the JC case, we see that Eq.~\ref{eq:Qsumeq} forms an invariant of the evolution.\label{fig:WWevol}}
\end{figure}

Our pair of atoms emitting into space from Eq.~\ref{eq:esdexample1} is then described by the state 
\begin{equation}\label{eq:WWinitialstate}
\ket{\phi}= \cos\theta \ket{g 0}_{(1)} \otimes \ket{g 0}_{(2)} + \sin\theta \ket{\Phi_t}_{(1)} \otimes \ket{\Phi_t}_{(2)}, 
\end{equation}
where subscripts $(1)$ and $(2)$ refer to atoms 1 and 2 respectively and their associated photon modes.  Tracing out either the atom or collective photon modes, results in ESD between atoms or photons~\cite{Lopez:08}, as can be seen in Fig.~\ref{fig:WWevol}.  Moreover, if we define $\ket{\Phi_t}$ as a moving basis state, then the entanglement between $\ket{\Phi_t}_{(1)}$ and $\ket{\Phi_t}_{(2)}$ is trivially constant in time.  Given the direct mapping, this solution (and subsequent results in section~\ref{sec:C4}) take the same form as the JC case and we therefore concentrate on our original JC example for most of the following work.  

\section{Hierarchies of geometric entanglement}\label{sec:HE}
While categorisation allows one to see that the system is evolving between inequivalent classes, it is possible to be more quantitative.  Using `hierarchies of geometric entanglement', one can also quantify the various contributions to the multipartite state~\cite{Wei:03,Blasone:08}.  The advantage of computing an entanglement hierarchy is that the various contributions can be quantified as a function of 2-, 3- and 4-partite entanglement as well as the dependence on how the system is partitioned.  

In order to compute an entanglement hierarchy, we define a set of $K$- separable states of an $N$-qubit system ($K\le N$).  In general, for values of $K<N$, there will be several different possible partitions where $Q_1 | Q_2 | \ldots |Q_K$ defines a particular partitioning of the system.  A familiar example for three qubits is that one can partition the system into one tripartite system ($K=N=3$) or three different arrangements of one qubit and one bipartite system ($K=2$).  We then take a general pure state $\ket{\Phi}$ which is defined by the set of all the $K$-separable states, $S_K(Q_1 | Q_2 | \ldots |Q_K)$, associated with a fixed partition, $K$ .  Calculating the overlap of such a state with the state in question $\ket{\psi}$ allows as to define the relative (partition dependent) geometric measure of entanglement  
\begin{equation}
E^{(K)}_{\rm{RGE}} (Q_1 | Q_2 | \ldots |Q_K) = 1-\Lambda_K^2(Q_1 | Q_2 | \ldots |Q_K),
\end{equation}
where 
\begin{equation}
\Lambda_K^2(Q_1 | Q_2 | \ldots |Q_K) = \max_{\ket{\Phi} \in S_K(Q_1 | Q_2 | \ldots |Q_K)} |\bra{\Phi} \psi \rangle|^2.
\end{equation}
The absolute (partition independent) geometric measure of entanglement is then defined by performing the maximisation over all of the possible partitions, $S_K$, for a given value of $K$, such that
\begin{equation}
E^{(K)}_{\rm{AGE}}(\ket{\psi}) = 1-\Lambda_K^2(\ket{\psi}),
\end{equation}
where 
\begin{equation}
\Lambda_K^2(\ket{\psi}) = \max_{\ket{\Phi} \in S_K} |\bra{\Phi} \psi \rangle|^2.
\end{equation}
The hierarchy of geometric entanglement is then defined by comparing the various contributions, $E_{\rm{AGE}}^{(2)} \le E_{\rm{AGE}}^{(3)} \le \ldots \le E_{\rm{AGE}}^{(N)}$.  In this hierarchy, $E_{\rm{AGE}}^{(N)}$ contains the total entanglement of the system, whereas $E_{\rm{AGE}}^{(N-1)}$ contains all except the bipartite.  The bipartite component is then computed by subtracting the outer level from the following one ($E_{\rm{AGE}}^{(N)}-E_{\rm{AGE}}^{(N-1)}$).  This recursion continues until $E_{\rm{AGE}}^{(2)}$, which measures the N-partite entanglement of the system.  As an example, a four qubit state with $E_{\rm{AGE}}^{(2)} = E_{\rm{AGE}}^{(3)} = E_{\rm{AGE}}^{(4)}$ contains only 4-partitie entanglement, whereas a state with $E_{\rm{AGE}}^{(2)} = E_{\rm{AGE}}^{(3)} < E_{\rm{AGE}}^{(4)}$ contains 4- and 2-partitie entanglement.  More details on the computation and interpretation of entanglement hierarchies can be found in Ref.~\onlinecite{Blasone:08}.

In this and subsequent sections, we use the notation $A_1 P_1 | A_2 P_2$ to indicate a partition which groups each atom and photon mode together, whereas $A_1 A_2 | P_1 P_2$ defines a partition grouping the two atoms together and the two photon modes together, forming separate atom-atom and photon-photon subsystems.  Throughout, it is also assumed that the state of the system is given by Eq.~(\ref{eq:JCsolution}) and therefore only a function of $\theta$ and $t$.

\begin{figure} [tb!]
\centering{\includegraphics[width=8cm]{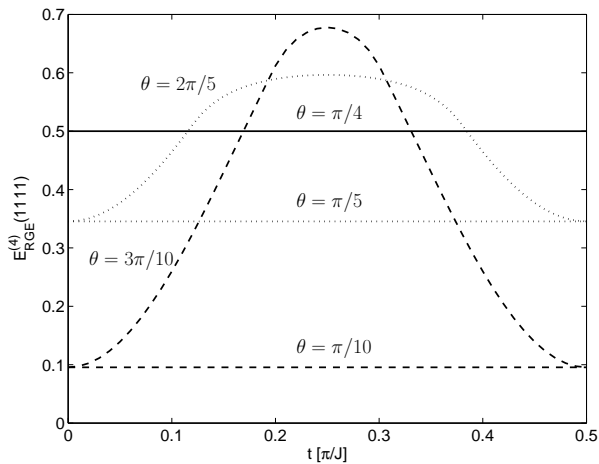}} 
\caption{Time evolution of $E_{\rm{AGE}}^{(4)}(1111)$ as a function of $\theta$ showing the asymmetry either side of $\theta = \pi/4$.  For values of $\theta > \pi/4$, the action of the JC interaction is to convert the initial bipartite entanglement into more general multipartite correlations. \label{fig:E1111evol}}
\end{figure}

Using entanglement hierarchies, we can make several important observations.  First, we plot the value of $E_{\rm{AGE}}^{(4)}(\ket{\psi(\theta,t)})$ for different values of $\theta$, see Fig.~\ref{fig:E1111evol}.  We immediately see the asymmetry with respect to $\theta$, where $E_{\rm{AGE}}^{(4)}(\ket{\psi(\theta,t)})$ is a constant as a function of time for $\theta \le \pi/4$.  For $\theta > \pi/4$ we see the total entanglement (4-, 3- and 2-partite) increase and then decrease as a function of time, corresponding to the JC interaction taking the initial bipartite entanglement and converting it to more complicated entangled states.  The asymmetry with respect to $\theta$ corresponds exactly with the asymmetry of ESD with respect to $\theta$.

\begin{figure} [tb!]
\centering{\includegraphics[width=8cm]{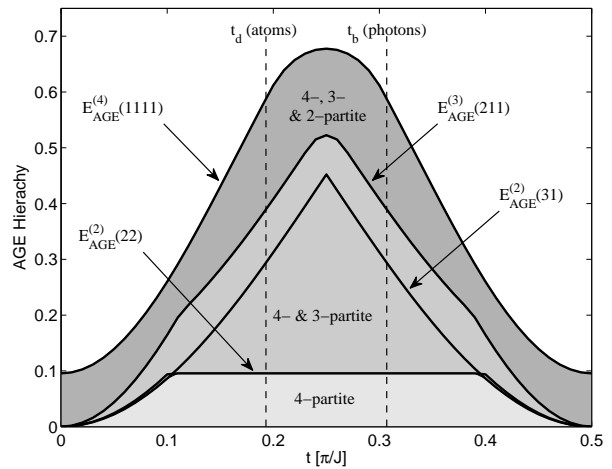}} 
\caption{Absolute Geometric Entanglement Hierarchy as a function of time for $\theta=2\pi/5$.  During the evolution, we see the amount of 2- and 3- partite entanglement increase, while the amount of 4-partite entanglement reaches a maximum equal to the initial amount of bipartite entanglement.  The point of entanglement sudden death (birth) for atoms (photons) is labelled $t_d$ ($t_b$) for this choice of $\theta$ but does not correspond to any special point of interest in the entanglement hierarchy.\label{fig:AGHE}}
\end{figure}

In Fig.~\ref{fig:AGHE}, we investigate the ESD region more precisely by plotting the various values of the absolute geometric entanglement as a function of time, for $\theta = 2\pi/5$.  We see the conversion of bipartite entanglement into 4-partite entanglement as well as the generation of a large fraction of 2- and 3-partite entanglement.  The total amount of available 4-partitie entanglement is limited by the amount of bipartite entanglement initially available, as one might expect.

\begin{figure} [tb!]
\centering{\includegraphics[width=8cm]{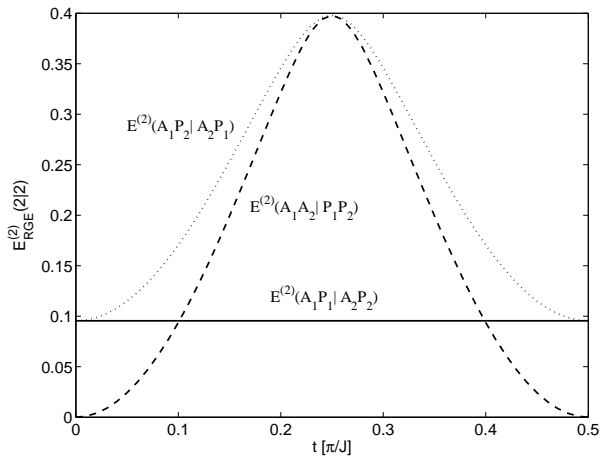}} 
\caption{Relative Geometric Entanglement for the partitions, $E_{\rm{RGE}}^{(2)}(2|2)$, which group the system into pairs.  We see that for the partition which seperates the atom--photon pairs, the entanglement is constant, as expected.  For other partitions, we see the cyclic increase and decrease in entanglement associated with action of the Jaynes--Cummings evolution between atom and photon.  In this figure, as with the previous examples $\theta=2\pi/5$, which is well within the ESD region.\label{fig:RGHE22}}
\end{figure}

In order to understand partition dependent effects, we plot the relative geometric entanglement $E_{\rm{RGE}}^{(2)}(\ket{\psi(2 \pi/5,t)})$ for three inequivalent partitions in Fig.~\ref{fig:RGHE22}.  The geometric entanglement associated with the partition which groups each atom--photon pair, $A_1 P_1 | A_2 P_2$, is constant.  This is consistent with the observation that this is a closed system with no interaction terms between the atoms or the photons.  In contrast, the other partitions show a marked increase during the JC oscillation.

It is important to note, while geometric entanglement hierarchies aids our understanding of the evolution of the system, at no point does either the absolute nor the relative geometric entanglement reach zero with a discontinuous derivative (i.e.\ displays ESD).  This suggest that concurrence is somehow `special' in this regard.

\section{Quantum correlations}
In fact, what is special about concurrence is that it measures separability of the system rather than the extent of correlations.  As a mixed state of two qubits can contain zero entanglement but nonzero quantum correlations (as typified by the Werner states~\cite{Werner:89, Luo:08}), we must also consider the role of quantum correlations.  In order to measure the quantum correlations of the system, we use quantum discord~\cite{Ollivier:01, Henderson:01,Luo:08}, $\mathcal{Q}(\rho)$, is a measure of the quantum correlations between two parties, whose combined state is given by $\rho$.  

\begin{figure} [tb!]
\centering{\includegraphics[width=8cm]{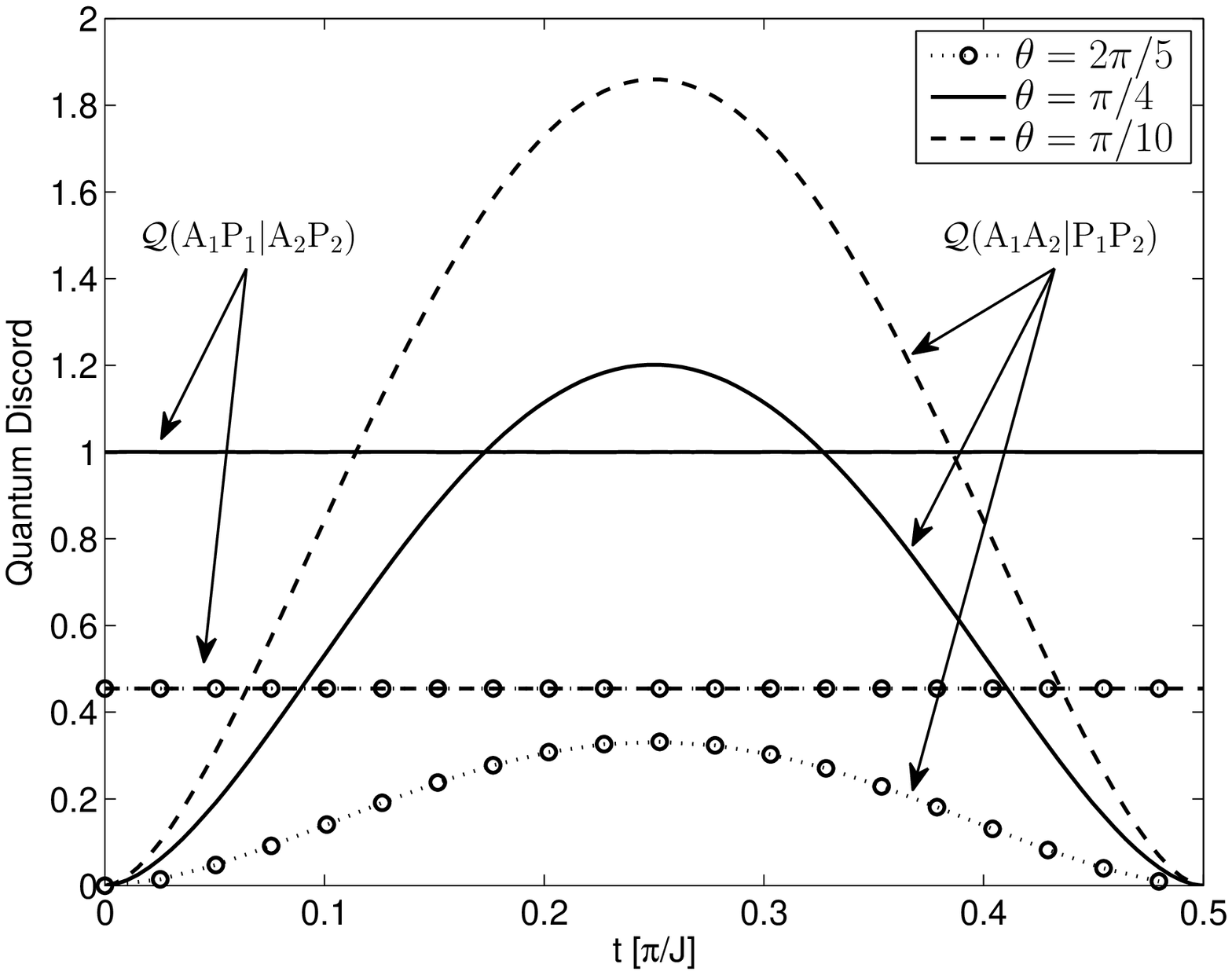}} 
\caption{Quantum discord as a function of time, computed for three state angles and two different partitions.  For the partition $\rm{A_1 P_1} | \rm{A_2 P_2}$ we see the discord does not change with time and is purely a function of the state angle $\theta$.  The partition $\rm{A_1 A_2} | \rm{P_1 P_2}$ captures the creation and then removal of correlations between atoms and photons.\label{fig:Discord}}
\end{figure}

For the JC system, it is natural to compute quantum discord for two in-equivalent partitions, one consisting of two atoms and two photons ($\rm{A_1 A_2} | \rm{P_1 P_2}$),  the other comprised of atom-photon pairs ($\rm{A_1 P_1} | \rm{A_2 P_2}$).  In Fig.~\ref{fig:Discord}, we plot the quantum discord for both these partitions for several different values of $\theta$.  For the partition $\rm{A_1 P_1} | \rm{A_2 P_2}$ we see the discord does not change with time and is purely a function of the state angle $\theta$.  It reaches a maximum of $\mathcal{Q}(\rm{A_1 P_1} | \rm{A_2 P_2})=1$ (one ebit) at $\theta = \pi/4$, where this correspond to one Bell state shared between the atom-photon pairs.  The discord is also symmetric about $\theta=\pi/4$, as we expect as there is no interaction which crosses this partition. The quantum discord is measuring the initial quantum correlations between the atoms, which are then smoothly transfered to the photons.  This case corresponds exactly to the `trivial' partitioning discussed earlier.

The partition $\rm{A_1 A_2} | \rm{P_1 P_2}$ captures the creation and then removal of correlations between atoms and photons.  As the photons have no initial population (and the atoms have no population after one half cycle) for this choice of partition, the quantum discord is zero at the extrema.  The quantum correlations reach a maximum halfway through the cycle, the value of which depends on $\theta$.  These are the correlations directly generated by the Jaynes--Cummings interaction and depend accordingly on $\theta$, varying from a zero (no correlation between atom and photon) to a maximum of $\mathcal{Q}(\rm{A_1 A_2} | \rm{P_1 P_2})=2$ when $\theta=\pi/2$.  In this limit, there is no initial quantum correlation or entanglement between the atoms and the two JC pairs evolve independently, producing two independent bell pairs, resulting in one ebit of entropy each. 

Computing the quantum discord tracks the evolution of the correlations in the system.  Even within an ESD region where the system is separable, there are still nonzero quantum correlations between the subcomponents.  The results from the Geometric Hierarchies can also be interpreted as measuring the total quantum correlations of the system, not just the entanglement~\footnote{The use of the term \emph{entanglement} hierarchy is therefore rather confusing but this convention is maintained here for continuity.}, and therefore do not drop to zero with a discontinuous derivative.  Furthermore, the ESD region can be directly identified with a region of state space where the state of the system is separable but still posses quantum correlations, which in turn is intimately linked to the concept of subsystem partitioning.

\section{Concurrence and invariants}\label{sec:C4}
We now return to using concurrence~\cite{Wootters:98} as a measure of entanglement which has both a pure state and mixed state definition.  The concurrence of a pure state $\ket{\Psi}$ is given by $C(\Psi) = |\langle \Psi \ket{\tilde{\Psi}}|$ where $\ket{\tilde{\Psi}}$ is the `spin-flipped' state $\ket{\tilde{\Psi}} = \sigma_y \otimes \sigma_y \ket{\Psi^*}$.  For a mixed state, $C(\rho) = \max [0,Q(\rho)]$ where $Q(\rho) = \sqrt{\lambda_1}- \sqrt{\lambda_2}- \sqrt{\lambda_3}- \sqrt{\lambda_4}$ is an auxiliary function of the eigenvalues of the spin-flipped density matrix $\rho \sigma_y \otimes \sigma_y \rho^* \sigma_y \otimes \sigma_y$, where both the conjugation and $\sigma_y$ are defined in the basis of interest.  As has been previously noted~\cite{Yu:07, Yu:09}, the region of ESD coincides with $Q(\rho)<0$ for the reduced density matrix of the atoms (or photons).  It is this region in which the value of the auxiliary function is reflecting the quantum correlations of the system, even though the system is separable.

We now consider the four-particle generalization of concurrence (or 4-tangle) as a quantitative entanglement measure~\cite{Wong:01,Coffman:00}.  This measure is constructed in an analogous way to bipartite concurrence, where $\ket{\tilde{\Psi}} = \sigma_y^{\otimes n}\ket{\Psi^*}$ for a $n$-qubit state.  For clarity, we will define $C_{ij}(\Psi)$ as the \emph{bipartite} concurrence between the $i$th and $j$th degree-of-freedom while tracing over the others, whereas $C_{4}(\Psi)$ is the four-particle concurrence of the entire atom--photon system.  In our JC example, the four-particle concurrence is
\begin{equation}
C_{4}[\psi(\theta, t)] = \frac{\sin^2\theta \sin^2 (2 J t)}{2}.
\end{equation}
 
Taking the time dependent state, $\ket{\psi(\theta, t)}$ for $\theta \le \pi/4$ and calculating the concurrence of the atoms and photons, $C_{\rm{AA}}(t)$ and $C_{\rm{PP}}(t)$ respectively, we find
\begin{eqnarray}\label{eq:sumeq}
C_{\rm{AA}}[\psi(\theta,0)] & = & C_{\rm{AA}}[\psi(\theta, t)]+ C_{\rm{PP}}[\psi(\theta, t)] + C_{4}[\psi(\theta, t)]  \nonumber \\
& = & \sin(2 \theta), \quad \theta \le \pi/4
\end{eqnarray}
valid for all time, $t$.  This suggests that the sum of the concurrence shared between the atoms, the photons and four-particle concurrence equals the initial concurrence in the system, implying that the `missing' entanglement is indeed four-particle entanglement.  This effect is illustrated graphically in Fig.~\ref{fig:jcevol}(a) where the bipartite concurrences, four-particle concurrence and Eq.~(\ref{eq:sumeq}) are plotted.  It is tempting to conclude that in Eq.~(\ref{eq:sumeq}) we have found an entanglement invariant but a little further calculation shows that this is not the case.  
In fact, it is exactly the regime that displays complete ESD ($\theta>\pi/4$) for which this relation is not valid.

The solution (as was hinted at by previous authors~\cite{FrancaSantos:06,Yu:07,Yu:09}) is to use the auxiliary function $Q_{\rm{AA}}[\rho(t)]$, rather than $C_{\rm{AA}}[\rho(t)]$, as this function has a nontrivial negative component in the ESD region, resulting from the quantum correlations.   Performing the necessary algebra, we find that indeed
\begin{eqnarray}\label{eq:Qsumeq}
\Sigma[\theta, t]  & = & Q_{\rm{AA}}[\psi(\theta, t)]+Q_{\rm{PP}}[\psi(\theta, t)] + C_{4}[\psi(\theta, t)] \nonumber \\
& = & \sin(2 \theta) \nonumber \\
& = & Q_{\rm{AA}}[\psi(\theta, t=0)], 
\end{eqnarray}
for all $t$ \emph{and} $\theta$, where we trace over unwanted qubits as required~\footnote{Very similar results to these are obtained if total variance~\cite{Klyachko:06, Klyachko:07} is used as an entanglement measure.}.  This supports the interpretation of the interplay between bipartite entanglement and multipartite entanglement, although the quantitative prediction of ESD is  unique to concurrence (and similar measures such as negativity) as no such behaviour is seen within the geometric entanglement hierarchies or quantum discord.  We also note that this invariant result is in direct contrast to that given in the erratum to Ref.~\onlinecite{Sainz:07}.  In that work, they chose a partitioning which groups the atom--photon pairs, resulting in trivially constant entanglement between the pairs, as we also saw in section~\ref{sec:HE} when computing $E_{\rm{RGE}}^{(2)}(A_1 P_1 | A_2 P_2)$.



\section{`Sudden' death via partitioning}
This dependance of ESD results on the partitioning of the system should come as no surprise, as the very notion of quantum entanglement is directly linked with the tensor product structure and the concept of partitioning~\cite{Zanardi:04}. To illustrate this, we return once more to the example of atoms spontaneously emitting into the vacuum.  Using WW theory, we are free to define our  boundaries between system and environment arbitrarily and therefore we can use this to explore the correspondence between time evolution (with fixed partitions) and movement of the system partition (at fixed time).  Taking the state from Eq.~\ref{eq:WWstate}, we redefine it into atom, collective state and environmental modes,
\begin{eqnarray}	
\ket{\Phi_t} & = & \xi(t)\ket{e}\ket{\mathbf{\overline{0}}} + \sum_{k\in K}^N \lambda_k(t) \ket{g} \ket{1_k} + \sum_{k\notin K}^N \lambda_k(t) \ket{g} \ket{1_k} \nonumber \\
& = &  \xi(t)\ket{e}\ket{0}\ket{0} + \chi'(t)\ket{g}\ket{\gamma_K}\ket{0}  \nonumber \\
&     &+  \sqrt{1-|\xi(t)|^2-|\chi'(t)|^2}\ket{g}\ket{0}\ket{\rm{env}}
\end{eqnarray}
where $k \in K$ is the set of modes that we consider part of a collective mode $\ket{\gamma_K}$ which can be entangled with either the atom or the other photon.  The remaining photon modes, $k \notin K$, are to be considered environmental modes and will be traced over.  Taking our example of two atoms in the state given by Eq.~(\ref{eq:WWinitialstate}), we consider our `system' to consist of only those photons belonging to set $K_{(1)}$ for atom one and $K_{(2)}$ for atom two.  For clarity of presentation, we will also take the long time limit such that
\begin{equation}\label{eq:WWinfstate}
\ket{\Phi_\infty} \approx \ket{g} \otimes \left(\chi'(\infty)\ket{\gamma}\ket{0} +  \sqrt{1-|\chi'(\infty)|^2}\ket{0}\ket{\rm{env}}\right),
\end{equation}
and $\chi'(\infty)$ is real.  The state $\ket{\Phi_K}=\chi'(\infty)\ket{\gamma}\ket{0} +  \sqrt{1-|\chi'(\infty)|^2}\ket{0}\ket{\rm{env}}$ is then analogous to a pair of coupled two-state systems, in direct correspondence with the atom--photon pairs (Eq.~\ref{eq:WWstate}) of the JC model in the single excitation limit.  In this case, the probability amplitudes are no longer time dependent but depend on choice of partitioning $k \in K$.
Setting $\chi'(\infty)_{(1)}=\chi'(\infty)_{(2)}=\chi'$, we find the reduced density matrix of the collective states $\ket{\Phi_{K_{(1)}}}\otimes\ket{\Phi_{K_{(2)}}}$ is of the $X$-form~\cite{Lopez:08}.  To observe entanglement between collective photon modes (given a fragile state), the system must satisfy
\begin{eqnarray}\label{eq:sumovermodes}
|\chi'|^2 & = & \sum_{k \in K_1,K_2} |\lambda_k(\infty)|_{(1)}^2|\lambda_k(\infty)|_{(2)}^2 \nonumber \\
& \ge & 1-\cot \theta, \quad \pi/4 \le \theta \le \pi/2.
\end{eqnarray}
which is not a function of time, but of partitioning of the system/environment.  This partitioning is parameterized by $|\chi'|^2$, the probability of observing the system in a collective mode (as opposed to an environmental mode).  As we include more states in the environmental modes, rather than the collective modes, eventually a sharp threshold is reached and entanglement can no longer be supported between these collective modes (given a fragile initial state).

To obtain a physical interpretation, we consider the output spectrum from a spontaneously emitting atom in the far field.  The spectrum is given by~\cite{Scully:97}
\begin{equation}
S(\nu) = \frac{\Gamma}{\pi} \frac{1}{(\mathcal{E}-\nu)^2+\Gamma^2}
\end{equation}
where integrating over all frequencies equals 1.  The probability of emitting a photon of frequency $\mathcal{E}-\Delta \nu \le \nu \le \mathcal{E} + \Delta \nu$ is given by the definite integral
\begin{equation}\label{eq:detectionprob}
p_{2\Delta \nu} = \int_{\mathcal{E}-\Delta \nu}^{\mathcal{E}+\Delta \nu} S(\nu) d \nu = \frac{2}{\pi} \arctan\left( \frac{\Delta \nu}{\Gamma} \right).
\end{equation}
If we take the modes $K_{(1)}$ and $K_{(2)}$ to have frequencies centred about the atom frequency $\mathcal{E}$ with width $\pm \Delta \nu$, this corresponds to an effective bandwidth of interest ($2 \Delta \nu$). 
Eq.~(\ref{eq:detectionprob}) and Eq.~(\ref{eq:sumovermodes}) lead to the following inequality,
\begin{equation}
\frac{\Delta \nu}{\Gamma} \ge \tan\left[ \frac{\pi}{2}(1-\cot \theta) \right], \quad \pi/4 \le \theta \le \pi/2,
\end{equation}
as the condition required to allow for entanglement between the two collective modes $\ket{\gamma_{K_{(1)}}}$ and $\ket{\gamma_{K_{(2)}}}$.  This makes physical sense, as the faster the decoherence, the more broad the spectral response and the more `classical' the emitted light.  
What is most interesting is that this process has a finite cutoff in partitioning, where entanglement goes abruptly to zero.  This is consistent with the interpretation of ESD resulting from Hamiltonian evolution between states with different entanglement visibilities, given a certain partitioning which is fixed in time.

\section{Conclusion}
Rather than considering entanglement sudden death (or birth) as a purely decoherence-induced effect, it is useful conceptually to treat it as a result of Hamiltonian evolution between entanglement classes of the larger system.  It is only when this larger system is traced over that the visibility of the multipartitie entanglement to \emph{bipartite} entanglement measures results in entanglement sudden death for the \emph{reduced bipartite system}.  This process is a direct consequence of the in-equivalence of entanglement and quantum correlations as a measure of the `quantumness' of a state.  

Using these concepts, we have shown the correspondence between the two most common examples of ESD as well as the fact that ESD can occur as a function of the choice of system partitioning, independent of time.  Furthermore, considering the multipartite generalization of concurrence allows us to define an entanglement invariant for both examples.  In contrast, the hierarchy of geometric entanglement and quantum discord provide a more general measure of the quantum correlations of the system and therefore do not predict ESD.  The phenomenon of ESD depends directly on both how the system is partitioned and whether one is interested in quantum entanglement or simply quantum correlations.

\section*{Acknowledgments}
The author thanks T. Tilma for suggesting the use of quantum discord and acknowledges useful discussions with S. Huelga, A. Osterloh, M. Terra Cunha, C. Hadley, A. Stephens and N. Oxtoby.  This work was completed with the support of the Alexander von Humboldt Foundation.  

\bibliography{esd}

\end{document}